# HANDY: A Hybrid Association Rules Mining Approach for Network Layer Discovery of Services for Mobile Ad hoc Network

Noman Islam, Zubair A. Shaikh, Aqeel-ur-Rehman, Muhammad Shahab Siddiqui

**Noman Islam** (Corresponding Author)

*National University of Computer and Emerging Sciences, Karachi, Pakistan*

Email: noman.islam@nuedu.pk

Zubair A. Shaikh

*National University of Computer and Emerging Sciences, Karachi, Pakistan*

Aqeel-ur-Rehman

*Hamdard University, Karachi, Pakistan*

Muhammad Shahab Siddiqui

*Hamdard University, Karachi, Pakistan*

**Abstract**

Mobile Ad hoc Network (MANET) is an infrastructure-less network formed between a set of mobile nodes. The discovery of services in MANET is a challenging job due to the unique properties of network. In this paper, a novel service discovery framework called Hybrid Association Rules Based Network Layer Discovery of Services for Ad hoc Networks (HANDY) has been proposed. HANDY provides three major research contributions. At first, it adopts a cross-layer optimized design for discovery of services that is based on simultaneous discovery of services and corresponding routes. Secondly, it provides a multi-level ontology-based approach to describe the services. This resolves the issue of semantic interoperability among the service consumers in a scalable fashion. Finally, to further optimize the performance of the discovery process, HANDY recommends exploiting the inherent associations present among the services. These associations are used in two ways. First, periodic service advertisements are performed based on these associations. In addition, when a response of a service discovery request is generated, correlated services are also attached with the response. The proposed service discovery scheme has been implemented in JIST/SWANS simulator. The results demonstrate that the proposed modifications give rise to improvement in hit ratio of the service consumers and latency of discovery process.

# 1 Introduction

Pervasive computing enables the technology to be inconspicuously weaved into daily lives [1] providing improved experience to end-user via context-aware execution of services in transparent manner. However, this integration leads to a host of issues arising due to ad hoc networking, nodes mobility and heterogeneity etc. The solution to these problems lies in the abstraction of various types of resources and device capabilities in the form of composable services called Service Oriented Architecture (SOA). These services are exposed via some interfaces (using XML or JSON etc). This abstraction helps in addressing the heterogeneous and dynamic temperament of the network.

Generally, a service is qualified by its description $D$ (like the input and output interfaces, functional description) and the QoS attributes $Q$ etc. Services are of different forms and complexities. A service can be software (image processing algorithm), hardware (printer/scanner), data (nearest restaurant information) or other type of resources (video) etc. Irrespective of the type of services, a service discovery mechanism is indispensable for the consumption of services. *Service discovery* is defined as the process of locating a service provider in the vicinity of service consumer based on the preferences (e.g. the description and QoS attributes) specified by the service consumer. Suppose $s_{ij}$ denotes the service $j$ located at node $i$ and $S_i$ denotes the set of services available at a particular node $i$ on the network i.e. $S_i = \{s_{i1}, s_{i2}, \ldots\}$, $S$ represents the set of services available on the network i.e., $S = \bigcup_{0 \le i \le n} S_i$ with $n$ denoting the number of nodes in the network and $s$ is the service requested by a service consumer. Then, the service discovery process can be mathematically defined as follows [35]:

$$discovery(s) = \{i | (0 \le i \le n) \bigwedge \exists j : (0 \le j \le S_i \ and \ s.D \approx s_{ij}.D \ and \ s.Q \le s_{ij}.Q)\} \quad (1)$$

Here, the symbol ≈ denotes the match operator i.e. the left hand side matches the right hand side (syntactically or semantically). This paper addresses the problem of service discovery in mobile ad hoc network (MANET). *MANET* is defined as an infrastructure-less, multi-hopped communication network created among a set of mobile nodes that are usually connected by volatile wireless links. Service discovery in MANET presents various challenges as compared to traditional networks. These challenges are due to their limited resources, mobility of nodes and heterogeneous nature etc. In view of these challenges, this paper presents a novel service discovery framework for MANET (HANDY) based on our initial concept outlined in [2-5]. As we will see in the next section (literature review), the proposed work is one of the pioneer approaches based on association rules mining and network layer service discovery. This paper piles up our past work in the form of a framework (as presented in section 3). It extends the work by proposing an associative service advertisement module. It also introduces essential formalism, protocol details and algorithms. The assumptions related to the framework have been validated in section 4. The framework is critically evaluated in section 5. The paper concludes with brief discussion on future work.

# 2 Literature Review

The problem of service discovery has been extensively researched both in conventional systems as well ad hoc networks. WS-Discovery is service discovery protocol for local networks. It was originally proposed by BEA Systems, Canon, Intel, Microsoft, and WebMethods and later on adopted by OASIS [6]. The protocol is based on multicast mechanisms and uses SOAP as the underlying mechanism for communication. XRDS [7] is an XML based protocol for discovery of metadata about a resource. It has been used for discovery of authentication services and context providers. [8] provides a DNS based approach for discovery of RESTful web services. Wang [9] proposes a scheme where data mining algorithms are applied over the usage log for optimization of service discovery operation.

The service discovery approaches proposed for conventional systems are not suitable for ad hoc networks, because they can't cope with the mobile and dynamic conditions of the network. Below, we discuss some of the schemes proposed for ad hoc networks.

### 2.1 Directory-based and directory-less approaches

Based on the service discovery architectures, we can classify the approaches as directory-based and directory-less approaches. In directory based approaches, service providers register their services to one or more directory servers that can be looked up by a service consumer to acquire information about the services. Examples of such approaches include JINI [10], Salutation [11], [12] and [13] etc. The directory based approaches are reliant on a set of specialized nodes called directory servers that can be predefined or elected at run time. These specialized servers might lead to performance bottlenecks, security and failure issues etc. To circumvent these issues, directory-less approaches can be used. In such approaches, every node maintains the service information locally in the form of a service table that can be queried on request to satisfy the discovery request. Examples of directory-less approaches are Microsoft UPnP [14], IBM DeapSpace [15], Bluetooth Service Discovery protocol [16] and Allia [17]. Besides the two approaches mentioned above, there are some service discovery protocols that can work in directory-based as well as directory-less fashion. For instance, Service Location Protocol [18] proposed by IETF works in hybrid fashion.

### 2.2 Cross-layer Service Discovery

Unlike the conventional approach that focuses on the architectural issues of service discovery, another aspect of service discovery which has been immensely stressed in the recent proposals is cross-layer discovery of services. A cross-layer solution works by integrating service discovery process with other layers. In [19-22], various ideas were presented for integration of service discovery with different routing protocols (AODV, DSR, ODMRP and OLSR etc.). In [23], the authors have used the Ad hoc On Demand Distance Vector Routing protocol for discovery of service in MANET. The authors propose to supplement the AODV header with several new fields such that the routing protocol can be used for discovery of services. The advantage of this maneuvering is that both service discovery and route discovery process is executed together thus minimizing the service discovery latency. In [24], a cross-layer service discovery scheme has been presented based on Ant-Colony optimization. The author proposes swarm routing based on collaboration among forward and backwards exploration agents analogous to ants.

### 2.3 Exploiting Data Mining for Service Discovery

Another distinctive direction of research on service discovery is the exploitation of the relationships present among the services for improving various performance parameters of discovery process. In this direction, we reported the earliest work in [4]. It is hypothesized that one can predict future discovery requests of a service consumer and responses of a discovery request can be attached with details of these predicted service discovery requests. Through experiments, it has been shown that if discovery requests made in a service session are correlated, performance gain can be observed by this piggybacking process. By applying FP-Growth mining algorithm on the service sessions to establish correlation patterns, anticipated discovery requests are then determined based on these correlation patterns. Bhole et al. [25] extends this approach further by using Apriori algorithm for finding out correlation patterns and reports improvement in discovery response time. In [26], data mining has been proposed for discovery of composite web services for client server systems. The Multilevel Association Rules Mining is applied to web service usage log to discover relationships among services and based on these relationships, suggestions are offered to the users. Moreover, the sequential pattern mining is used to order the set of web services.

### 2.4 Semantic Service Discovery

Semantic service discovery enables discovery of services based on the underlying meaning of the services. This enables interoperability among heterogonous nodes in the network. Some existing service discovery approaches have been extended to provide semantics to the services. For example, DReggie [27] extends JINI by plugging in a component that performs semantic matching of the services. The matching component uses prolog based reasoner to perform the semantic lookup of service. Several new approaches have been proposed for semantic service discovery. GSD [28] uses DAML based ontology to group the services. The grouping of nodes is used for intelligent forwarding of service requests on the network. Konark [29] is a middleware based on SOAP and XML to semantic discovery of services in a device independent manner on a peer-to-peer ad hoc network.

### 2.5 Other Approaches

Besides the research directions outline above, various other approaches have also been presented in literature. In [30], an analogy has been developed between the MANET and the electrostatic field. On the basis of this analogy, the authors have proposed a mathematical model for service discovery in MANET. In [31], a service discovery approach is presented based on ZeroConf protocol. The algorithm triggers its service advertisement process upon any topological changes in the network, hence minimizing the expenses of the discovery process. In [32], a service discovery scheme is proposed in which a set of mobile nodes fetches the services' information at regular intervals from its surrounding nodes. These mobile nodes can then entertain the queries of service consumers by providing the relevant information about the queried services.

This paper proposes a novel service discovery framework (HANDY). Compared with currently available frameworks, it provides three major contributions. It works by integrating with underlying routing protocol (AODV/DSR) and provides simultaneous discovery of services and corresponding routes. The cross-layer design leads to minimum resource utilization and also provides robustness against the spatiotemporal variations in the network. Second contribution provided by HANDY is an ontological approach to service representation. There are few ontology based schemes (DReggie [27], GSD [28] and Konark [29]) already available in this direction. However, HANDY proposes a multi-level ontology documents for service representation. It recommends maintaining portion of the global schema locally and discovering rests of the schema in on-demand style. This approach ensures interoperability among the nodes in a scalable way. Finally, HANDY recommends using the correlation pattern among the services to optimize different aspects of the discovery process. These associations are used for proactive advertising of services and piggybacking with response of a discovery request.

## 3. Proposed Service Discovery Scheme (HANDY)

HANDY stands for a *H*ybrid *A*ssociation Rules Mining based Approach for *N*etwork Layer *D*iscover*Y* of Services in Mobile Ad hoc Network. Fig 1 shows the block diagram of HANDY describing the main components running at every node on the network. It comprises of various core components for service discovery and service representation as well as supporting components for data mining and logging purposes.

The *logging* component maintains the history of past service requests in the form of service sessions. Each of the service requests is logged in a circular linked list called log database. The *log database* $L = \{H_1,H_2,H_3,\ldots\}$ maintains the requested services' information via a linked list of service sessions. We define a *service session* as a sequence of requests $H_i = \{r_1,r_2, \ldots r_h\}$ such that time difference between issuance of any two consecutive requests is below a threshold $\xi$ i.e. $\forall i<h: r_i+1.t - r_i.t<\xi$., where $r_i.t$ denotes the time at which the request was issued. The *association rules mining* component applies the FP-Growth mining algorithm on the log database to compute the relationships among the services. The results of this association rules mining algorithm are stored in a list called mining results. The mining results are used by the service discovery component. The service discovery component works in a hybrid style for discovery of services. It periodically broadcasts services (based on the results of mining) information about services (HANDY-P) and when required can discover the services reactively on the network (HANDY-R). The *service catalog* keeps the record of the services available on the network. There are three tables maintained for this purpose: *service table*, *routing table* and the *ontology table*. In addition an ontology database is also maintained. The service table maintains the information about the services hosted by the node or by some other nodes as shown in Table 1a. The information includes service ID, service name, provider, input and output interface of the service, list of ontology required for understanding the service and QoS attributes related to the service. The routing table maintains the routing information for propagating the routing and service request to other nodes on the network. Table 1b shows the routing table maintained by every node on the network. Table 1c shows the meta-information maintained by every node about the ontology. This includes the name of ontology, its URI, author and the ontology which is extended by this ontology.

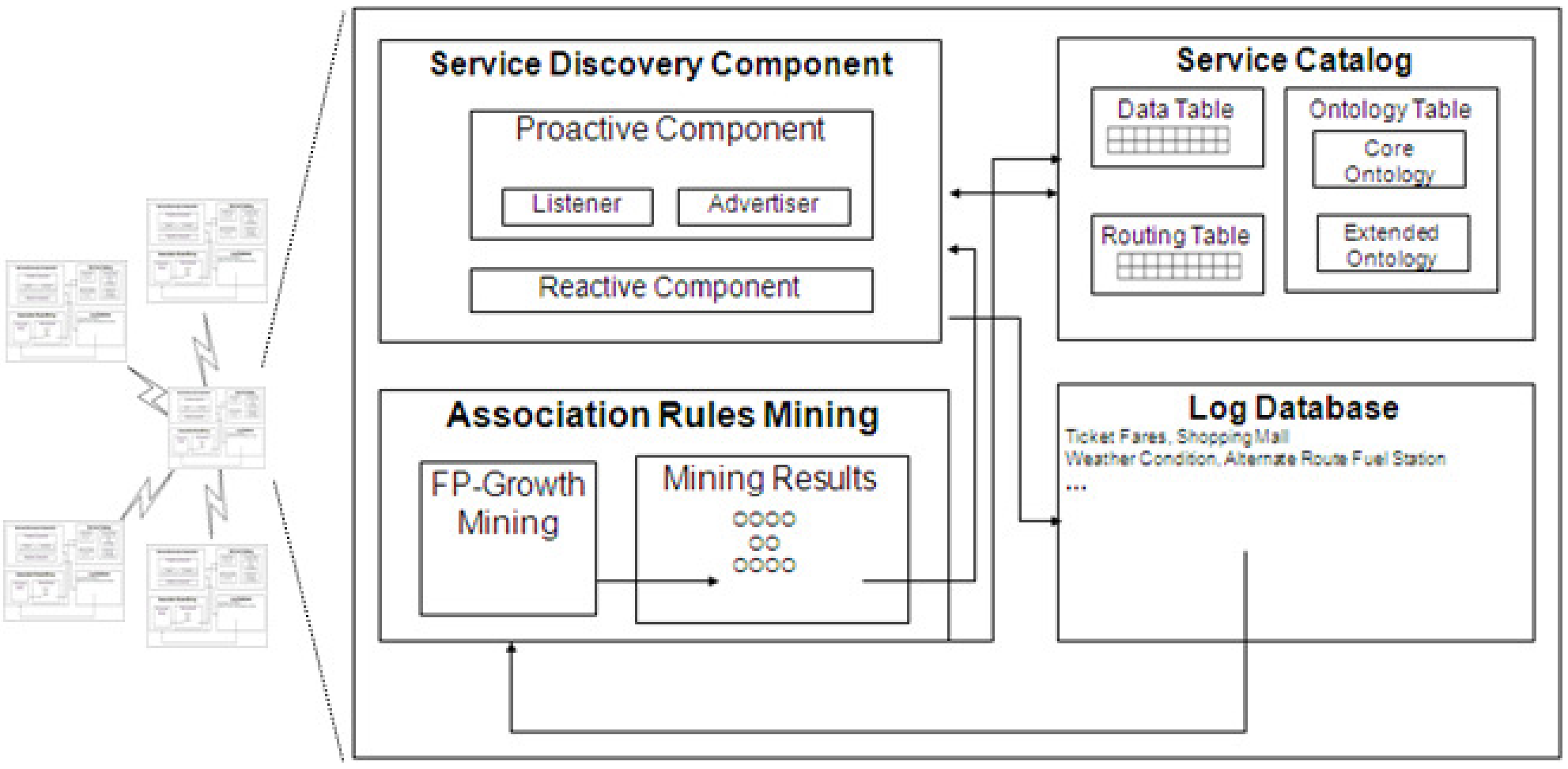

**Figure 1: Block diagram describing the components of HANDY**

**Table 1: Service, Routing and Ontology tables maintained by HANDY nodes**

| **Field** | **Description** |
|---|---|
| Service ID | The unique identifier for the service |
| Service Name | The name of the service |
| Service Provider | The identity of the node providing the service |
| Input Interface | The input interface to call the service |
| Output Interface | The output interface of the service |
| Ontology List | The list of ontology documents required to understand services |
| QoS | The QoS information associated with the service |

a) Service Table

| **Field** | **Description** |
|---|---|
| Sequence Number | Determines the freshness of the routing information |
| Destination | The destination for which routing information is maintained |
| Hop Count | The number of hop counts from this node to the destination |
| Next Node | The next node to whom packet will be forwarded for the particular destination |
| Status | Status of the route |
| Precursors | The upstream nodes of a particular route |

b) Routing Table

| **Field** | **Description** |
|---|---|
| Name | The name of the ontology |
| URI | The URI of the ontology |
| Author | The name of the author who created the ontology |
| Base Ontology | The URI of the ontology extended by this ontology |
| Ontology | A pointer to the location from which this ontology can be downloaded |

c) Ontology Table

The ontology component maintains the semantic information required to understand the meaning of the services. This ensures the interoperability among the nodes. To ensure the scalability of nodes, two-level ontology is maintained. All the nodes maintain a *core ontology* that describes the essential concepts of any MANET application. The core ontology is extended by the *ext ontology* documents for specialized applications. The ext ontology is not maintained for every node. A node can download the ext ontology from other nodes of the network by invoking the HANDY-R module. This on-demand ontology management approach relieves the nodes from processing and storage burden.

### 3.1 Service Discovery Component

The service discovery component comprises of two sub-components: a proactive component (HANDY-P) and a reactive component (HANDY-R). The former is used for proactive maintenance of services need in apriori fashion. The later is used for discovery of services from other nodes of the network in on-demand fashion.

#### *3.1.1 HANDY-P*

The proactive component is composed of an advertiser and a listener module. Listing 1 shows the pseudo code for the advertiser component. The *advertiser* sends associative service advertisements *SADV* at regular intervals (*p, 2p,3p ….*) to neighboring nodes. The advertisements contain details about a set of services that are correlated to each other. The *listener* component receives these advertisements and saved in local service table. If $m$ is the length of an advertisement, then $A_{it} = \{a_1,a_2,a_3,\ldots a_m\}$ represents the advertisement by a node $i$ at time $t$. Suppose $\rho(x,y)$ represents the correlation coefficient between two variables, then for a particular service $a$, to become a part of advertisement it must be correlated to all the services that are part of advertisement i.e.$\forall_{j<m}\, \rho\left(a, a_j\right) > \varepsilon$. Here, $\varepsilon$ represents a threshold value such that correlation between any two services must be greater than this threshold to be advertised together. For any neighbor $\eta \in \eta_{it}$, let $\oplus$ denotes the operator that is used by node $\eta$ to update its service records, then the advertisements $A_{it}$ is cached by listener component of $p$ as follows [35]:

$$S_n = S_n \oplus \forall_{a \in A_{it}}\, a \tag{2}$$

**Listing 1: HANDY-P**

**[HANDY Advertiser Component]**

```
Procedure Advertiser()
{
      While  (True)
      {
          If ( mode  = "Correlated")
          {
             s =  getLeastRecentlyBroadcastedService();
           broadCastItems.add(s);
           broadCastItems = minedResults.GetRelated(s);

          }
          Else
          {
           broadCastItems  =
                      GetNLeastRecentlyBroadcastedServices();
          }
          routes = RoutingTable.GetRoutes
                      (broadcastItems);
          Send(broadCastItems,routes);
          wait (sleep_time);
      }
}
```

```
Procedure Listener()
{

  While(True)
  {
    m = ReceivePacket ();
    If(m.Type = "SADV")
    {
       Save (m broadCastItems, m.routes);
    }
  }
}
```

#### *3.1.2 HANDY-R*

The reactive component HANDY-R is activated on-demand when a node $i$ don't find information about a requested service $s$ in its local service table. Listing 2 shows the pseudo code for the operation of HANDY-R. It floats a request on the network for discovery of the desired service. The discovery request is propagated using the underlying routing protocol. Upon activation, the reactive component compiles a service request SREQ and sends to its neighboring nodes $\eta_{it}$. Any node $\eta$ receiving this request will check its local table for the availability of requested information. The service matching component (section 3.2) ensures that a service is available and its QoS complies with the QoS specified in the request. In case of a service match, the node will construct a reply SREP and sends this reply back to the source node *i*. Otherwise, the node will propagate the service request information ahead to its neighboring nodes. The propagation of SREQ by intermediate nodes continues until the request reaches to a node that has the information about the requested service or the TTL value of the SREQ expires. The nodes receiving the SREP will update their service table accordingly. For further details of the AODV and DSR based discovery, please refer to [2].

**Listing 2: HANDY-R**

```
[HANDY Reactive Service Discovery Component]

Procedure HANDY_R()
{
      While  (True)
      {
        p = ReceivePacket();

        If (p is SREQ)
        {
          Service s = MatchService(p);

          if (s <> Empty)
          {
             SendUpStream (GenerateReply(p, s));
          }

          Else
          {
             Propagate(p);
          }
        }

        If (p is SREP)
        {
          If (p.destination = host.IP)
          {
             UpdateServiceTable();
             ConsumeService(p);

          }
          Else
          {
             UpdateServiceTable();
             SendUpstream(p);
          }
        }
      }
}

Procedure Propagate()
{
  If ( NetworkLayer.RoutingProtcol = AODV)
  {
     CreateReverseEntry();
     destination += GetAdjacentNodes (p);
  }
  Else If ( NetworkLayer.RoutingProtcol = DSR)
  {
     p.HopsTraversed +=  host.IP;
    destination += GetAdjacentNodes (p);
  }
  Else If ( NetworkLayer.RoutingProtcol = TORA)
  {
     [processing for TORA]
  }
  [similarly for other routing protocols]
  …

  Send (p, destination);
}

Procedure SendUpStream()
{
    If ( NetworkLayer.RoutingProtcol = AODV)
    {
       destination = ReverseEntry.destination;
       CreateForwardEntry();
    }
    Else If ( NetworkLayer.RoutingProtcol = DSR)
    {
        destination =  Head (p.HopsTraversed);

    }
    Else If ( NetworkLayer.RoutingProtcol = TORA)
    {
       [processing for TORA]
    }
    [similarly for other routing protocol]
    …
    Send (p, destination);
}
```

### *3.1.3 Using Correlation Patterns to Optimize the Reactive Discovery Process*

While generating the SREP by any node $\eta$, the node also takes the advantage of correlation among the services to attach responses of expected future request with the current response. This is based on the assumption (as validated in section 4) that those services that were used together in past are expected to be used together in future as well. In other words, if $s$ is the requested service, then the reply generated by the node $\eta$ will contain information about the following services:

$$SREP = \{s \bigcup \forall_j \ \rho(i,j) > \ni\} \tag{3}$$

**SADV message**

| Field | Details |
|---|---|
| Message Id | Unique Id of every message |
| Type of Message | AODV in this case |
| Length | Length of the message |
| Correlated Service List. Each list entry includes: | List of Correlated Services |
| a) Service Details | Details of a service i.e. service id, name, provider, input interface, output interface, ontology list and QoS |
| b) Ontology Details | Details of the ontology required to understand the service i.e. name, URI, author, base ontology and ontology pointer |
| b) Routing Details | Details of corresponding routing entry i.e. sequence #, destination, hop count, next node, status and Precursors |

**SREQ message**

| Field | Details |
|---|---|
| Message Id | Unique Id of every message |
| Session Id | The Session Id of the request |
| Type of Message | SREQ in this case |
| Length | Length of the message |
| Requested Service | Details of the requested service |
| Ontology List | List of ontology required to understand the service |
| Routing Headers | The essential routing headers for propagation of service requests. This include details as mentioned in table 1b as well as an hops traversed field in case of DSR |

**SREP message**

| Field | Details |
|---|---|
| Message Id | Unique Id of every message |
| Session Id | The Session Id of the request |
| Type of Message | SREP in this case |
| Length | Length of the message |
| Requested Service | Details of the requested service |
| Ontology List | List of ontology required to understand the service |
| Routing Headers | The essential routing headers for propagation of service requests. This include details as mentioned in table 1b as well as an hops traversed field in case of DSR |
| List of Correlated Service | List of service correlated with requested service along with corresponding route and ontology details |

**Figure 2: Type of Messages used by HANDY**

### *3.1.4 Protocol Message Details*

Fig 2 shows various types of requests issued by the service discovery component. The advertisements contain details about the message like its ID, type, length as well as a list of services that are correlated. The correlated service list contains details about the service, corresponding routing details for each service and the list of ontology documents required to understand the service. The SREQ message includes message ID, type, session ID, length, details of requested service, routing headers, QoS details and list of ontology document required to understand the requests. In case of SREP, a field hops traversed is also maintained which is used during propagation of service replies back to the source. Similarly, the SREP includes message ID, type of message, session ID, length, provider details and ontology document list to understand the reply.

## 3.2 Semantic Component

MANET based applications suffer from an important issue of semantic non-interoperability. The problem arises among the hosts using different types of hardware/software components and belonging to variety of domains. These hosts can use proprietary data representation techniques leading to interoperability issues when communicating with each other. In addition, dealing with different domain may generate conflicts when using different terminologies for similar service. To resolve this issue, the nodes on the network are required to have a common schema/model describing the meaning of the terms used during exchange. The participating entities can thus consult the schema to reach to a common agreement about the terminologies used in their communication. One of the approaches to describe the semantic is software ontology. An ontology defines the meaning of terminologies, their attributes and relationships by using an ontology language [33]. However, the use of a unified schema is not possible in MANET due to its infrastructure-less nature. To address this problem, HANDY has adopted multi-level ontology approach as shown in Fig 3. HANDY comprise of two levels of ontology. The core level ontology is a generalized ontology containing essential concepts related to the domain and is available at every node. The ext ontology extends the core ontology for specialized applications. It is not necessarily available at every node and nodes can exchange these ontology documents when required.

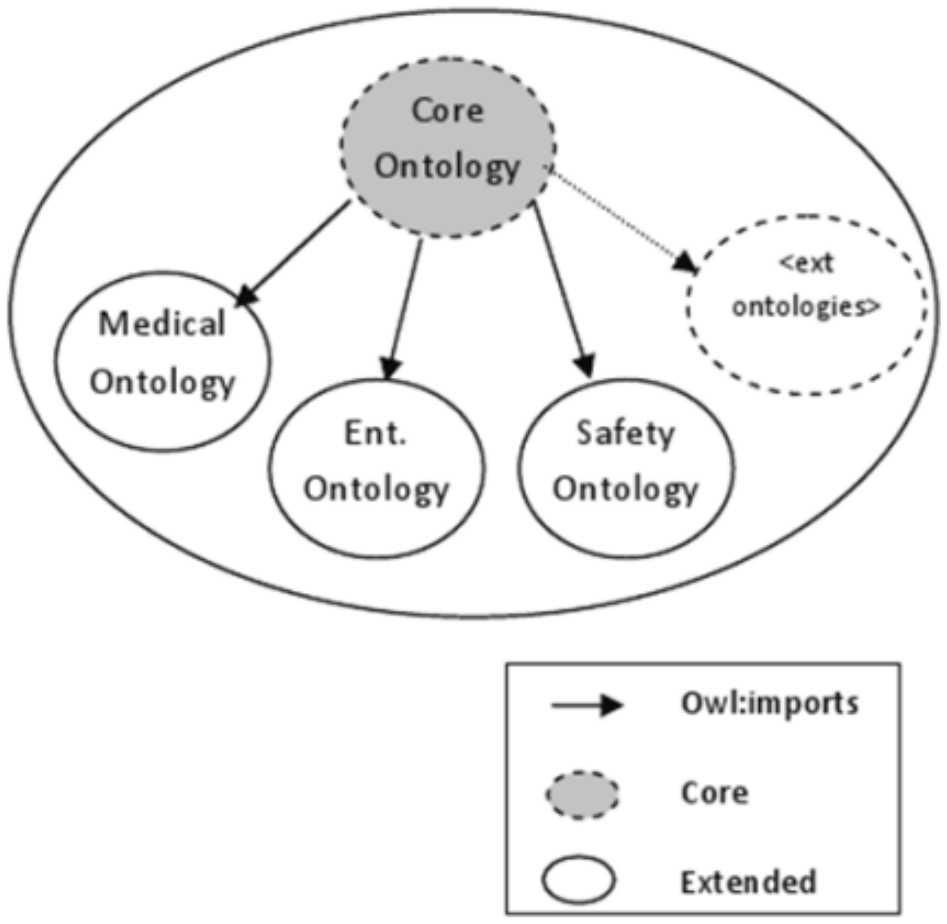

**Figure 3: HANDY's Multi-level Ontology Approach [5]**

### *3.2.1 SLAVE: Software Ontology for Ad hoc and Vehicular Network Applications*

As a case study for HANDY, we have developed general purpose software ontology for ad hoc and vehicular network applications (SLAVE). Fig 4 shows a sample view of proposed ontology in OWL format. The core ontology comprises of a set of core concepts needed for any application in the domain as shown in Fig 4a. This includes concepts related to a network node, person, place and related profiles etc. Three types of ext ontology for the domain of medical, safety and entertainment are proposed. The medical ontology describes the concepts related to medical applications. It includes details about doctor, health profile and disease etc. The safety ontology comprises of concepts useful for providing improved driving experience. Finally, the entertainment ontology embraces concepts like shopping places and items in the vicinity and their associated costs.

```
<owl:Class rdf:ID="Driver">
  <rdfs:subClassOf>
    <owl:Class rdf:ID="Person"/>
  </rdfs:subClassOf>
</owl:Class>
<owl:Class rdf:ID="Profile"/>
<owl:Class rdf:ID="VehicleProfile">
  <rdfs:subClassOf rdf:resource="#Profile"/>
</owl:Class>
<owl:Class rdf:ID="VANETNode">
  <rdfs:subClassOf>
    <owl:Class rdf:ID="Ad_hoc_Node"/>
  </rdfs:subClassOf>
</owl:Class>
<owl:Class rdf:ID="Place"/>
<owl:Class rdf:ID="DirvingProfile">
  <rdfs:subClassOf rdf:resource="#Profile"/>
</owl:Class>
<owl:Class rdf:ID="DrivingProfile">
  <rdfs:subClassOf rdf:resource="#Profile"/>
</owl:Class>
<owl:Class rdf:ID="ComputationalProfile">
  <rdfs:subClassOf rdf:resource="#Profile"/>
</owl:Class>
<owl:Class rdf:ID="SpatialProfile">
  <rdfs:subClassOf rdf:resource="#Profile"/>
</owl:Class>
<owl:ObjectProperty rdf:ID="hasComputationalProfile">
  <rdfs:domain rdf:resource="#Ad_hoc_Node"/>
  <rdfs:range rdf:resource="#ComputationalProfile"/>
</owl:ObjectProperty>
<owl:ObjectProperty rdf:ID="hasDrivingProfile">
  <rdfs:range rdf:resource="#DirvingProfile"/>
  <rdfs:domain rdf:resource="#VANETNode"/>
</owl:ObjectProperty>
<owl:ObjectProperty rdf:ID="hasVehicleProfile">
  <rdfs:range rdf:resource="#VehicleProfile"/>
  <rdfs:domain rdf:resource="#VANETNode"/>
</owl:ObjectProperty>
<owl:ObjectProperty rdf:ID="hasSpatialProfile">
  <rdfs:range rdf:resource="#SpatialProfile"/>
  <rdfs:domain>
    <owl:Class>
      <owl:unionOf rdf:parseType="Collection">
        <owl:Class rdf:about="#Ad_hoc_Node"/>
        <owl:Class rdf:about="#Place"/>
      </owl:unionOf>
    </owl:Class>
  </rdfs:domain>
</owl:ObjectProperty>
<owl:DatatypeProperty rdf:ID="maxCycles">
  <rdfs:domain rdf:resource="#ComputationalProfile"/>
  <rdfs:range rdf:resource="#int"/>
</owl:DatatypeProperty>
<owl:DatatypeProperty rdf:ID="freeMemory">
  <rdfs:range rdf:resource="#int"/>
  <rdfs:domain rdf:resource="#Ad_hoc_Node"/>
</owl:DatatypeProperty>
<owl:DatatypeProperty rdf:ID="speed">
  <rdfs:domain rdf:resource="#Ad_hoc_Node"/>
  <rdfs:range rdf:resource="#int"/>
</owl:DatatypeProperty>
<owl:DatatypeProperty rdf:ID="y">
  <rdfs:range rdf:resource="#int"/>
  <rdfs:domain rdf:resource="#SpatialProfile"/>
</owl:DatatypeProperty>
<owl:DatatypeProperty rdf:ID="maxMemory">
  <rdfs:domain rdf:resource="#ComputationalProfile"/>
  <rdfs:range rdf:resource="#int"/>
</owl:DatatypeProperty>
<owl:DatatypeProperty rdf:ID="height">
  <rdfs:domain rdf:resource="#DrivingProfile"/>
  <rdfs:range rdf:resource="#int"/>
</owl:DatatypeProperty>
<owl:DatatypeProperty rdf:ID="gender">
  <rdfs:range rdf:resource="#string"/>
  <rdfs:domain rdf:resource="#Person"/>
</owl:DatatypeProperty>
<owl:DatatypeProperty rdf:ID="direction">
  <rdfs:range rdf:resource="#int"/>
  <rdfs:domain rdf:resource="#Ad_hoc_Node"/>
</owl:DatatypeProperty>
<owl:DatatypeProperty rdf:ID="x">
  <rdfs:range rdf:resource="#int"/>
  <rdfs:domain rdf:resource="#SpatialProfile"/>
</owl:DatatypeProperty>
<owl:DatatypeProperty rdf:ID="freeCycles">
  <rdfs:domain rdf:resource="#ComputationalProfile"/>
  <rdfs:range rdf:resource="#int"/>
</owl:DatatypeProperty>
<owl:DatatypeProperty rdf:ID="age">
  <rdfs:domain rdf:resource="#Person"/>
  <rdfs:range rdf:resource="#string"/>
</owl:DatatypeProperty>
<owl:DatatypeProperty rdf:ID="width">
  <rdfs:domain rdf:resource="#VehicleProfile"/>
  <rdfs:range rdf:resource="#int"/>
</owl:DatatypeProperty>
<owl:DatatypeProperty rdf:ID="name">
  <rdfs:domain rdf:resource="#Person"/>
  <rdfs:range rdf:resource="#string"/>
</owl:DatatypeProperty>
```

**Core Ontology**

```
<owl:Class rdf:ID="HealthProfile">
  <rdfs:subClassOf rdf:resource="#Profile"/>
</owl:Class>
<owl:Class rdf:ID="Hospital">
  <rdfs:subClassOf rdf:resource="#Place"/>
</owl:Class>
<owl:Class rdf:ID="Doctor">
  <rdfs:subClassOf rdf:resource="#Person"/>
</owl:Class>
<owl:Class rdf:ID="MedicalProfile">
  <rdfs:subClassOf rdf:resource="#Profile"/>
</owl:Class>
<owl:ObjectProperty rdf:ID="hasHealthProfile">
  <rdfs:range rdf:resource="#HealthProfile"/>
  <rdfs:domain rdf:resource="#Driver"/>
</owl:ObjectProperty>
<owl:ObjectProperty rdf:ID="hasDoctors">
  <rdfs:range rdf:resource="#Doctor"/>
  <rdfs:domain rdf:resource="#Hospital"/>
</owl:ObjectProperty>
<owl:ObjectProperty rdf:ID="hasSkillProfile">
  <rdfs:domain rdf:resource="#Doctor"/>
</owl:ObjectProperty>
<owl:DatatypeProperty rdf:ID="diseaseList">
  <rdfs:domain rdf:resource="#HealthProfile"/>
  <rdfs:range rdf:resource="#string"/>
</owl:DatatypeProperty>
<owl:DatatypeProperty rdf:ID="qualification">
  <rdfs:range rdf:resource="#string"/>
  <rdfs:domain rdf:resource="#MedicalProfile"/>
</owl:DatatypeProperty>
<owl:DatatypeProperty rdf:ID="name">
  <rdfs:domain rdf:resource="#Hospital"/>
  <rdfs:range rdf:resource="#string"/>
</owl:DatatypeProperty>
<owl:DatatypeProperty rdf:ID="specialization">
  <rdfs:range rdf:resource="#string"/>
  <rdfs:domain rdf:resource="#MedicalProfile"/>
</owl:DatatypeProperty>
```

**Medical Ontology**

```
<owl:Class rdf:ID="Status"/>
<owl:Class rdf:ID="TrafficStatus">
  <rdfs:subClassOf rdf:resource="#Status"/>
</owl:Class>
<owl:Class rdf:ID="WeatherStatus">
  <rdfs:subClassOf rdf:resource="#Status"/>
</owl:Class>
<owl:DatatypeProperty rdf:ID="statusId">
  <rdfs:range rdf:resource="#int"/>
  <rdfs:domain rdf:resource="#Status"/>
</owl:DatatypeProperty>
<owl:DatatypeProperty rdf:ID="statusDescription">
  <rdfs:range rdf:resource="#string"/>
  <rdfs:domain rdf:resource="#Status"/>
</owl:DatatypeProperty>
```

**Safety Ontology**

```
<owl:Class rdf:ID="ShoppingItem"/>
<owl:Class rdf:ID="ShoppingPlace">
  <rdfs:subClassOf rdf:resource="#Place"/>
</owl:Class>
<owl:ObjectProperty rdf:ID="hasShoppingItems">
  <rdfs:range rdf:resource="#ShoppingItem"/>
  <rdfs:domain rdf:resource="#ShoppingPlace"/>
</owl:ObjectProperty>
<owl:DatatypeProperty rdf:ID="name">
  <rdfs:range rdf:resource="#string"/>
  <rdfs:domain rdf:resource="#ShoppingPlace"/>
</owl:DatatypeProperty>
<owl:DatatypeProperty rdf:ID="itemname">
  <rdfs:domain rdf:resource="#ShoppingItem"/>
  <rdfs:range rdf:resource="#string"/>
</owl:DatatypeProperty>
<owl:DatatypeProperty rdf:ID="qty">
  <rdfs:range rdf:resource="#string"/>
  <rdfs:domain rdf:resource="#ShoppingItem"/>
</owl:DatatypeProperty>
<owl:DatatypeProperty rdf:ID="unitprice">
  <rdfs:domain rdf:resource="#ShoppingItem"/>
  <rdfs:range rdf:resource="#string"/>
</owl:DatatypeProperty>
```

**Entertainment Ontology**

**Figure 4: HANDY Multi-level Ontology Management Approach**

Listing 3 shows the working of HANDY's service matching component based on multi-level ontology management operation. It first searches in its ext ontology database for the availability of ontology required to understand the service request. In case of unavailability, it is searched by invoking the reactive discovery component and then stored in the ontology database. The component then initializes the inference model and queries for the desired service matching, the description and QoS specification.

**Listing 3: Semantic matching of services by HANDY**

```
Function Match (Message m) : Service
{
    required_ont = m.RequiredOntology;
    If ( extOntology.contains(required_ont) = False)
    {
        ontology = HANDY_R.discover(required_ont);
        ExtOntology.Store(ontology);
    }
    OntologyDatabase.DoInferencing();
    Query = "Select service_name, service_provider, … from Services where m.service.D ~Services.D and m.service.QoS
~ Services.QoS";
    Result = ExtOnt.Search (Query);
}
```

# 4. Assumptions and their validation

HANDY is based on the assumption that service discovery requests posed in a session are correlated to each other. We presented earlier work based on this assumption in [4,2]. In this section, we validate this basic assumption by analyzing the data of a large network setting. To establish the notion, an FP-Growth mining algorithm [34] has been applied to a dataset obtained from the HTTP requests issued for videos of YouTube by users in a campus over a period of time [35]. As we discussed earlier, services are of various forms. These services are basically abstraction around resources of various types. These resources can be a process dedicated for complex mathematical operation, specific hardware, an audio or a video resource etc. Hence, we believe that the requests for YouTube videos can be analyzed to discover the relationships among the services in a service discovery session.

The YouTube dataset is organized as an array of sessions (called as service sessions) with each session containing a set of the requested videos (called services). The results of the FP-Growth mining algorithm (with YouTube dataset as input) are compared with the results produced when a randomly generated dataset has been used as input to the mining algorithm. For the random dataset, a session H with length h is generated as follows:

$$H(i) = \begin{cases} random(0, |S|) & i = 0 \\ j : j = random(0, |S|) \wedge \forall_{k \leq i}\, j \neq H(k) & 1 \leq i \leq l \end{cases} \quad (4)$$

Fig 5 compares the number of frequent item sets produced by the mining algorithm when it is applied on the two datasets. A frequent item set of services is defined as a subset of services $F \subseteq S$ that are requested together repeatedly for a large number of times. Fig 5a shows various item sets (1,2,3,4 or more) produced when several number of services' (video) requests were considered for applying mining algorithm. Fig 5b shows the total number of frequent item sets generated when different numbers of services are considered. From Fig 5, we can see a clear distinction between the numbers of frequent item sets generated for the two datasets.

| | Services = 1000 | | Services = 1500 | | Services = 2000 | |
|---|---|---|---|---|---|---|
| | YouTube Data Set | Random Data Set | YouTube Data Set | Random Data Set | YouTube Data Set | Random Data Set |
| 1-Item Sets | 43 | 20 | 40 | 15 | 40 | 10 |
| 2-Item Sets | 34 | 0 | 34 | 0 | 32 | 0 |
| 3-Item Sets | 41 | 0 | 40 | 0 | 39 | 0 |
| 4 or above | 36 | 0 | 30 | 0 | 33 | 0 |

a) Number of n-Item Sets for the two datasets

| Services | YouTube Data Set | Random Data set |
|---|---|---|
| 50 | 119 | 0 |
| 100 | 156 | 10 |
| 500 | 177 | 35 |
| 1000 | 154 | 20 |
| 1500 | 144 | 15 |
| 2000 | 144 | 10 |
| Maximum | 177 | 35 |
| Minimum | 119 | 0 |
| Average | 149 | 15 |

b) Total Number of Frequent ItemSets for the two datasets

**Figure 5: Comparison of number of frequent item sets for YouTube and random data set**

Fig 6 also highlights this distinction by means of a bar diagram. It can be concluded that the YouTube dataset produces relatively higher number of item sets where as the random dataset produces low number of item sets which is definitely because of absence of any relationships among the services. The distribution of services is purely random in the second dataset, consequently producing small number of item sets. Based on the results, it can be concluded that the requests issued in a particular session are correlated; otherwise they would not have been used together for considerable number of times. We can also conclude that requests issued by consumers are not random at all and are influenced by the correlation among the services.

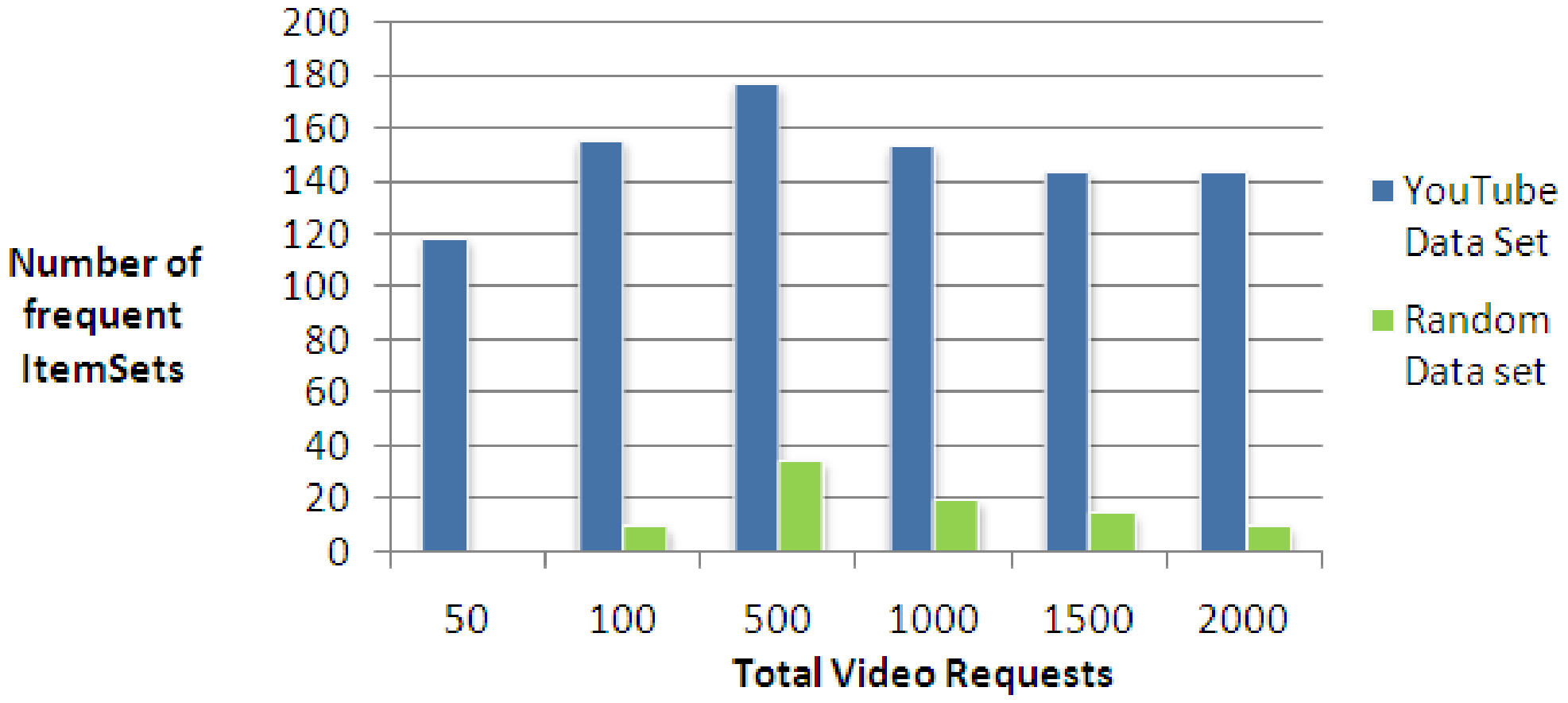

**Figure 6: Bar Diagram illustrating total number of frequent ItemSets for the two datasets**

# 5 Implementation Details

To analyze the performance of HANDY, a number of simulation based experiments have been conducted using the JIST/SWANS simulator [36]. Table 2 shows the parameters of simulation. Most of the parameters of the simulation were kept to default values of the JIST/SWANS simulator unless explicitly mentioned. For estimating the energy consumption of nodes,

Friedman's extensions have been proposed [37]. The dimension of the simulation field is 500 × 500, in which the nodes are randomly placed. The cache replacement strategy used to update service table is LRU. Two types of mobility models were used i.e. random way point and street mobility models. For ontology representation, we have used OWL language. The proposed ontology (as discussed in section 3.2) has been developed using Protégé [38]. For processing of ontology documents at runtime, JENA toolkit [39] has been selected. Three types of datasets were used to float discovery requests during simulation, as discussed below. These datasets are used to spawn service discovery sessions of varying characteristics.

**Table 2: Simulation Parameters**

| Parameters | Value |
|---|---|
| Simulation Area | 500 × 500 |
| Node Placement | Random |
| Simulation Time | 5000 |
| Resolution Time | 60 |
| Transport Protocol | UDP |
| Path Loss Model | Free Space |
| Mobility Models | Random Way Point and Street Mobility (Suffolk County Map) |
| Correlation Coefficient ($\varepsilon$) | 0.5 |
| CDG Threshold ($\eta$) | 0.5 |
| Caching | LRU |
| Data Representation | Data Identifiers |
| Default Node Speed | 1-20 |

**Random Dataset**

The random dataset is used to generate a service discovery sessions in which there are no correlation among discovery requests. The data set is generated based on Java Random class as follows:

$$H(0) = random(0, |S|) \qquad (5)$$
$$H(i) = random(0, |S|) \wedge \forall\ k \leq i : j \neq H(k)$$

**Correlated Dataset**

This dataset is generated such that the discovery requests are correlated. A random correlation matrix is first generated. Based on this correlation matrix, the service sessions are spawned. The details can be seen in [2,4].

**YouTube based Dataset**

Instead of using a random correlation matrix, a correlation matrix is generated as the output of Equation 4 that can be used to generate service sessions. Table 3 shows the correlation matrix used for generating sessions.

# 6 Results

This section analyzes the performance of various components of HANDY based on the impact of these components on latency and hit ratio of the discovery process. We will also analyze the energy consumption of nodes at different stages of simulation. We define *latency* as the time when a service discovery request was issued to the time its response was received by the consumer node. The *hit ratio* is defined as the ratio of the number of service requests satisfied locally to the total number of service requests issued. *Energy consumed* is the sum of energy consumed by nodes in sleeping, idle, receiving, transmitting and sensing modes.

**Table 3: Correlation Matrix based on YouTube video requests of a campus' users**

| | 1 | 2 | 3 | 4 | 5 | 6 | 7 | 8 | 9 | 10 | 11 | 12 | 13 | 14 | 15 | 16 | 17 | 18 | 19 | 20 | 21 | 22 | 23 | 24 | 25 |
|---|---|---|---|---|---|---|---|---|---|---|---|---|---|---|---|---|---|---|---|---|---|---|---|---|---|
| 1 | 0 | 0 | 0 | 0 | 0 | 0 | 0 | 0 | 0 | 0 | 0 | 0 | 0 | 0 | 0 | 0 | 0 | 0 | 0 | 0 | 0 | 0 | 0 | 0 | 0 |
| 2 | 0 | 0 | 0 | 0 | 0 | 0 | 0 | 0 | 0 | 0 | 0 | 0 | 0 | 0 | 0 | 0 | 0 | 0 | 0 | 0 | 0 | 0 | 0 | 0 | 0 |
| 3 | 0 | 0 | 0 | 1 | 1 | 0 | 1 | 0 | 0 | 0 | 0 | 0 | 0 | 0 | 0 | 0 | 0 | 0 | 0 | 0 | 0 | 0 | 0 | 0 | 0 |
| 4 | 0 | 0 | 1 | 0 | 1 | 0 | 1 | 1 | 0 | 0 | 0 | 0 | 0 | 0 | 0 | 0 | 0 | 0 | 0 | 0 | 0 | 0 | 0 | 0 | 0 |
| 5 | 0 | 0 | 1 | 1 | 0 | 0 | 1 | 1 | 0 | 0 | 0 | 0 | 0 | 0 | 0 | 0 | 0 | 0 | 0 | 0 | 0 | 0 | 0 | 0 | 0 |
| 6 | 0 | 0 | 0 | 0 | 0 | 0 | 1 | 1 | 0 | 1 | 0 | 0 | 0 | 0 | 0 | 0 | 0 | 0 | 0 | 0 | 0 | 0 | 0 | 0 | 0 |
| 7 | 0 | 0 | 1 | 1 | 1 | 1 | 0 | 1 | 0 | 0 | 0 | 0 | 0 | 0 | 0 | 0 | 0 | 0 | 0 | 0 | 0 | 0 | 0 | 0 | 0 |
| 8 | 0 | 0 | 0 | 1 | 1 | 1 | 1 | 0 | 0 | 0 | 0 | 0 | 0 | 0 | 0 | 0 | 0 | 0 | 0 | 0 | 0 | 0 | 0 | 0 | 0 |
| 9 | 0 | 0 | 0 | 0 | 0 | 0 | 0 | 0 | 0 | 0 | 0 | 0 | 0 | 0 | 0 | 0 | 0 | 0 | 0 | 0 | 0 | 0 | 0 | 0 | 0 |
| 10 | 0 | 0 | 0 | 0 | 0 | 1 | 0 | 0 | 0 | 0 | 0 | 0 | 0 | 0 | 0 | 0 | 0 | 0 | 0 | 0 | 0 | 0 | 0 | 0 | 0 |
| 11 | 0 | 0 | 0 | 0 | 0 | 0 | 0 | 0 | 0 | 0 | 0 | 0 | 0 | 0 | 0 | 0 | 0 | 0 | 0 | 0 | 0 | 0 | 0 | 0 | 0 |
| 12 | 0 | 0 | 0 | 0 | 0 | 0 | 0 | 0 | 0 | 0 | 0 | 0 | 0 | 0 | 0 | 0 | 0 | 0 | 0 | 0 | 0 | 0 | 0 | 0 | 0 |
| 13 | 0 | 0 | 0 | 0 | 0 | 0 | 0 | 0 | 0 | 0 | 0 | 0 | 0 | 1 | 1 | 0 | 1 | 0 | 0 | 0 | 0 | 0 | 0 | 0 | 0 |
| 14 | 0 | 0 | 0 | 0 | 0 | 0 | 0 | 0 | 0 | 0 | 0 | 0 | 1 | 0 | 0 | 0 | 1 | 1 | 0 | 0 | 0 | 0 | 0 | 0 | 0 |
| 15 | 0 | 0 | 0 | 0 | 0 | 0 | 0 | 0 | 0 | 0 | 0 | 0 | 1 | 0 | 0 | 0 | 1 | 1 | 0 | 0 | 0 | 0 | 0 | 0 | 0 |
| 16 | 0 | 0 | 0 | 0 | 0 | 0 | 0 | 0 | 0 | 0 | 0 | 0 | 0 | 0 | 0 | 0 | 0 | 1 | 0 | 1 | 0 | 0 | 0 | 0 | 0 |
| 17 | 0 | 0 | 0 | 0 | 0 | 0 | 0 | 0 | 0 | 0 | 0 | 0 | 1 | 1 | 1 | 0 | 0 | 0 | 0 | 0 | 0 | 0 | 0 | 0 | 0 |
| 18 | 0 | 0 | 0 | 0 | 0 | 0 | 0 | 0 | 0 | 0 | 0 | 0 | 0 | 1 | 1 | 1 | 0 | 0 | 0 | 0 | 0 | 0 | 0 | 0 | 0 |
| 19 | 0 | 0 | 0 | 0 | 0 | 0 | 0 | 0 | 0 | 0 | 0 | 0 | 0 | 0 | 0 | 0 | 0 | 0 | 0 | 0 | 0 | 0 | 0 | 0 | 0 |
| 20 | 0 | 0 | 0 | 0 | 0 | 0 | 0 | 0 | 0 | 0 | 0 | 0 | 0 | 0 | 0 | 1 | 0 | 0 | 0 | 0 | 0 | 0 | 0 | 0 | 0 |
| 21 | 0 | 0 | 0 | 0 | 0 | 0 | 0 | 0 | 0 | 0 | 0 | 0 | 0 | 0 | 0 | 0 | 0 | 0 | 0 | 0 | 0 | 0 | 0 | 0 | 0 |
| 22 | 0 | 0 | 0 | 0 | 0 | 0 | 0 | 0 | 0 | 0 | 0 | 0 | 0 | 0 | 0 | 0 | 0 | 0 | 0 | 0 | 0 | 0 | 0 | 0 | 0 |
| 23 | 0 | 0 | 0 | 0 | 0 | 0 | 0 | 0 | 0 | 0 | 0 | 0 | 0 | 0 | 0 | 0 | 0 | 0 | 0 | 0 | 0 | 0 | 0 | 0 | 0 |
| 24 | 0 | 0 | 0 | 0 | 0 | 0 | 0 | 0 | 0 | 0 | 0 | 0 | 0 | 0 | 0 | 0 | 0 | 0 | 0 | 0 | 0 | 0 | 0 | 0 | 0 |
| 25 | 0 | 0 | 0 | 0 | 0 | 0 | 0 | 0 | 0 | 0 | 0 | 0 | 0 | 0 | 0 | 0 | 0 | 0 | 0 | 0 | 0 | 0 | 0 | 0 | 0 |

### 6.1 Service Discovery Component

Fig 7 analyzes the improvement in latency obtained by incorporating various optimization techniques proposed by the HANDY service discovery component. By implementing the AODV based reactive service discovery component [23], an average latency of 22.5 sec is obtained. Further incorporation of advertisement modules gives rise to significant improvement in these parameters. The rationale behind these improvements is that the advertisement leads to the propagation of service (and route details in advance). Most of the service requests' details are thus available locally. By plug-in the piggybacking module, we can see that overall latency improves further. However, it is to be observed that the results remain almost same irrespective of the use of FP-Growth or Apriori algorithm [25].

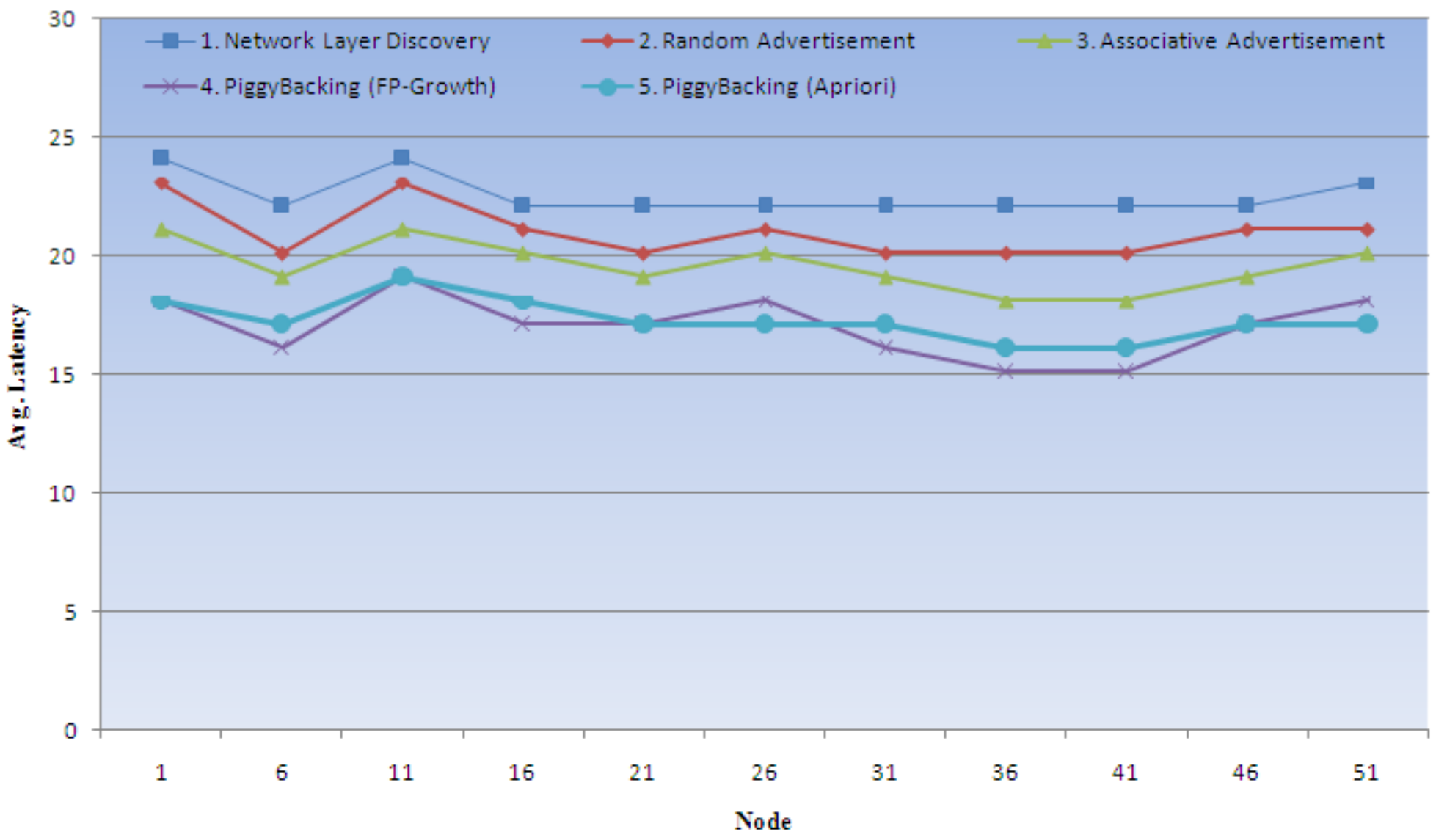

| | | | | | |
|---|---|---|---|---|---|
| Number of Nodes | 50 | Number of Services | 32 | Correlation Matrix | CDG |
| Routing Protocol | AODV | Mobility Model | Street Mobility | Cache Size | 8 |

**Figure 7: Latency Analysis of Service Discovery Component**

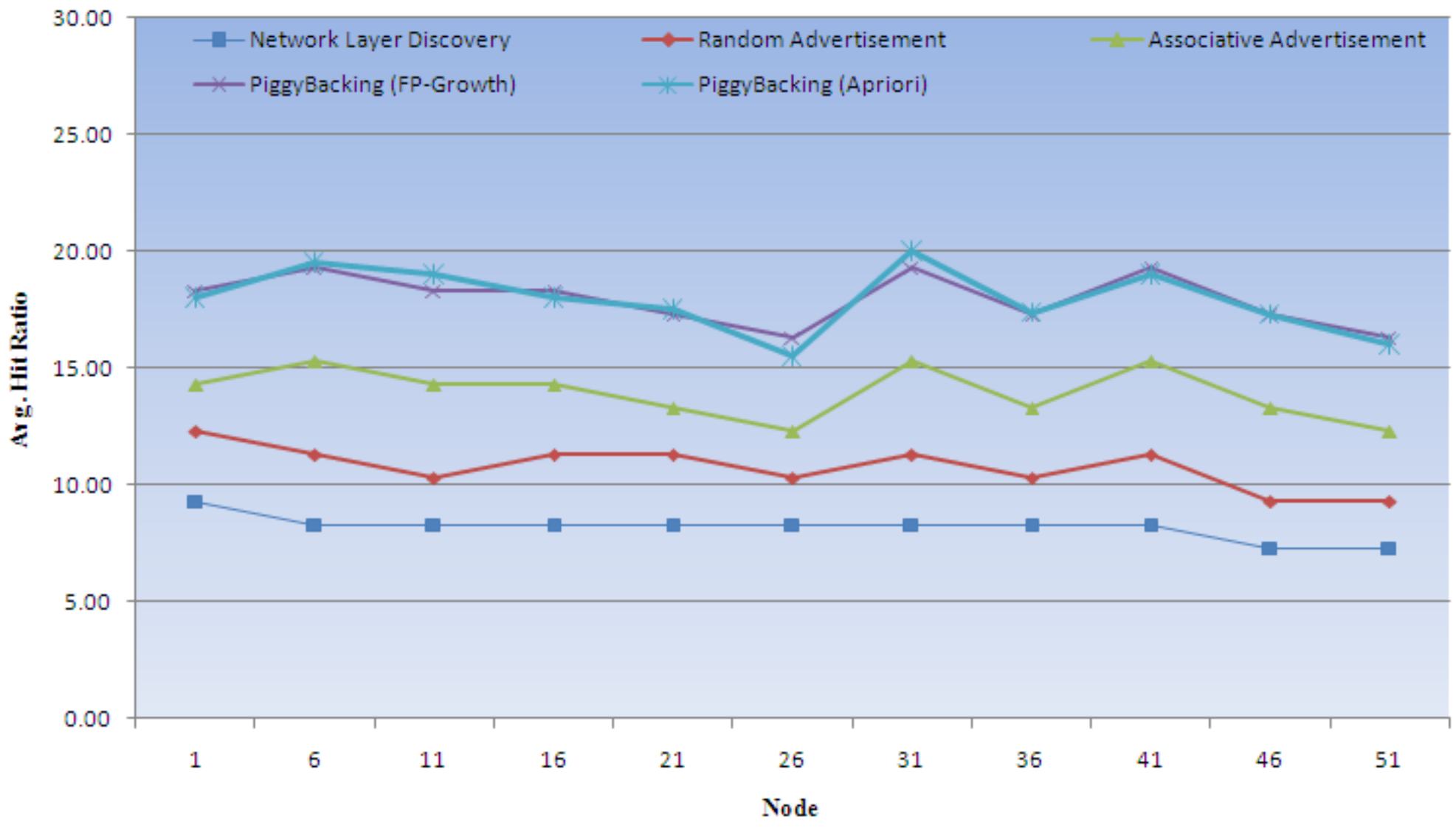

| | | | | | |
|---|---|---|---|---|---|
| Number of Nodes | 50 | Number of Services | 64 | Correlation Matrix | YouTube |
| Routing Protocol | DSR | Mobility Model | Street Mobility | Cache Size | 8 |

**Figure 8: Hit Ratio Analysis of Service Discovery Component**

Fig 8 investigates the hit ratio of the service discovery component. It is evident from the graph that the hit ratio improves with the use of service advertisements and piggybacking module. Fig 9 further analyzes the impact of different cache sizes and speed on hit ratio of the discovery component. As expected, the average hit ratio improves with the increase in cache size because of the availability of more slots to store service advertisements. What is more interesting is that the hit ratio also increases with the speed of nodes. The rise in speed enables nodes to come in interaction with each other and exchange services' requests and advertisements. With the nodes having more information about the services, intrinsic relationships can be accurately estimated.

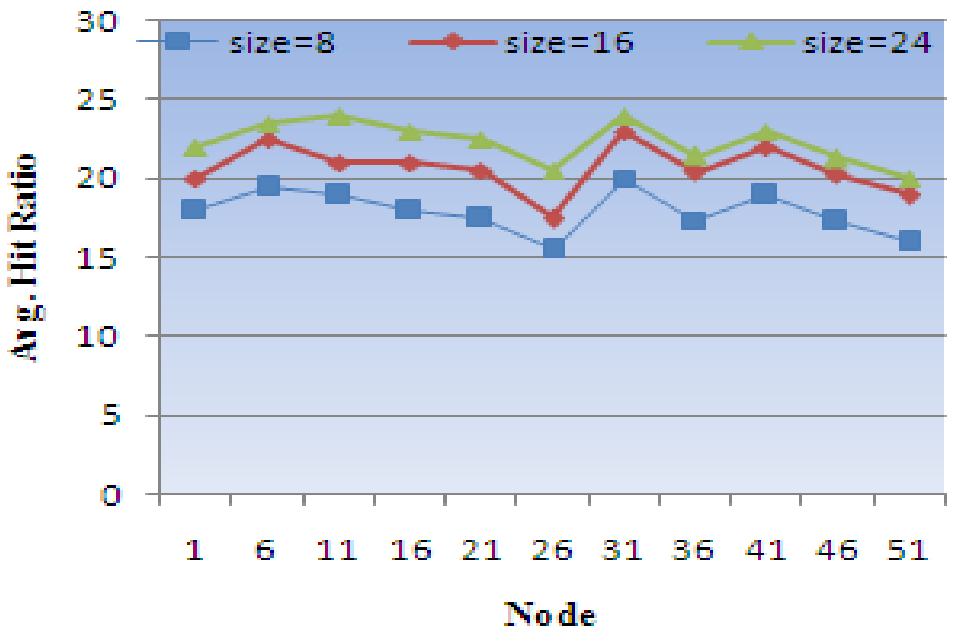

a) Analysis with different cache sizes

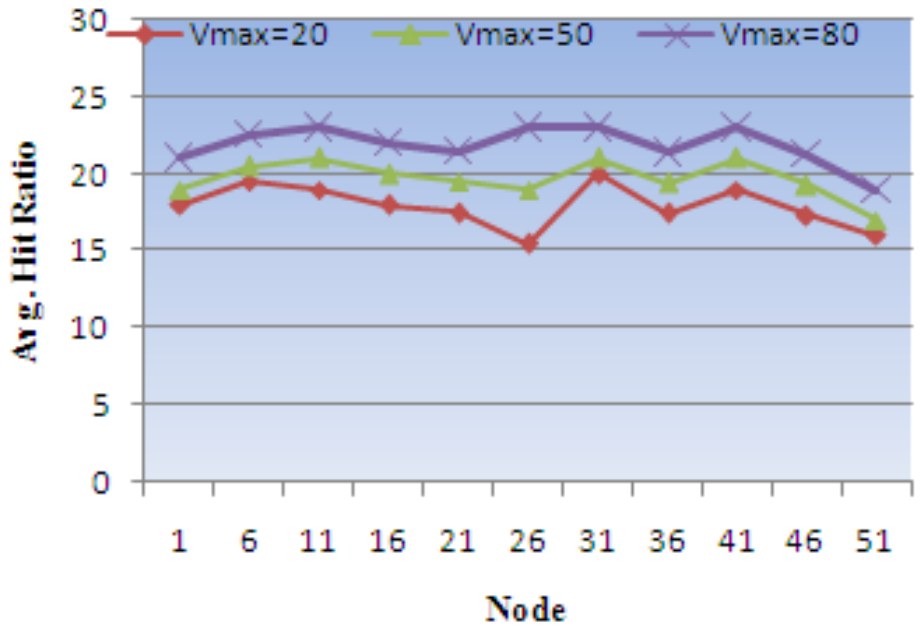

b) Analysis with different speeds

| | | | | | |
|---|---|---|---|---|---|
| Number of Nodes | 50 | Number of Services | 64 | Correlation Matrix | YouTube |
| Routing Protocol | DSR | Mobility Model | Random Waypoint | | |

**Figure 9: Hit Ratio Analysis with cache size and node speeds**

We close this section with the analysis of energy consumption at the nodes running the HANDY framework. Fig 10a shows average energy consumption of HANDY service discovery component with various numbers of nodes. Fig 10b further shows the energy consumption by individual components of nodes. The incorporation of advertisement module leads to more energy consumption. However, the associative service advertisement leads to almost same energy consumption as the random advertisement module. Similarly, the piggybacking module doesn't lead to any additional overhead in terms of energy consumption.

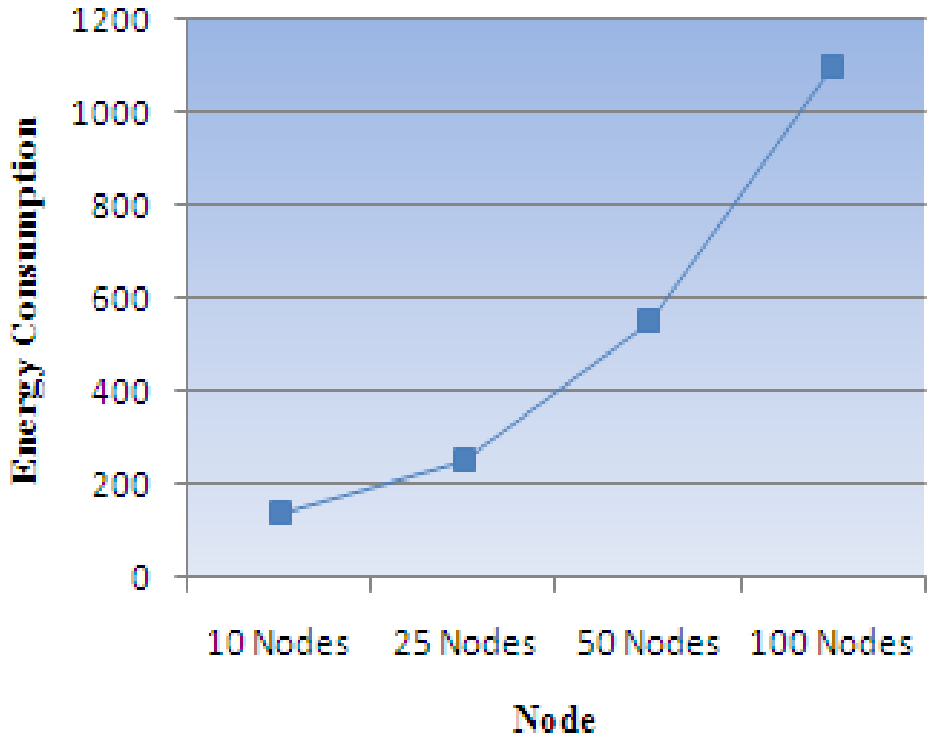

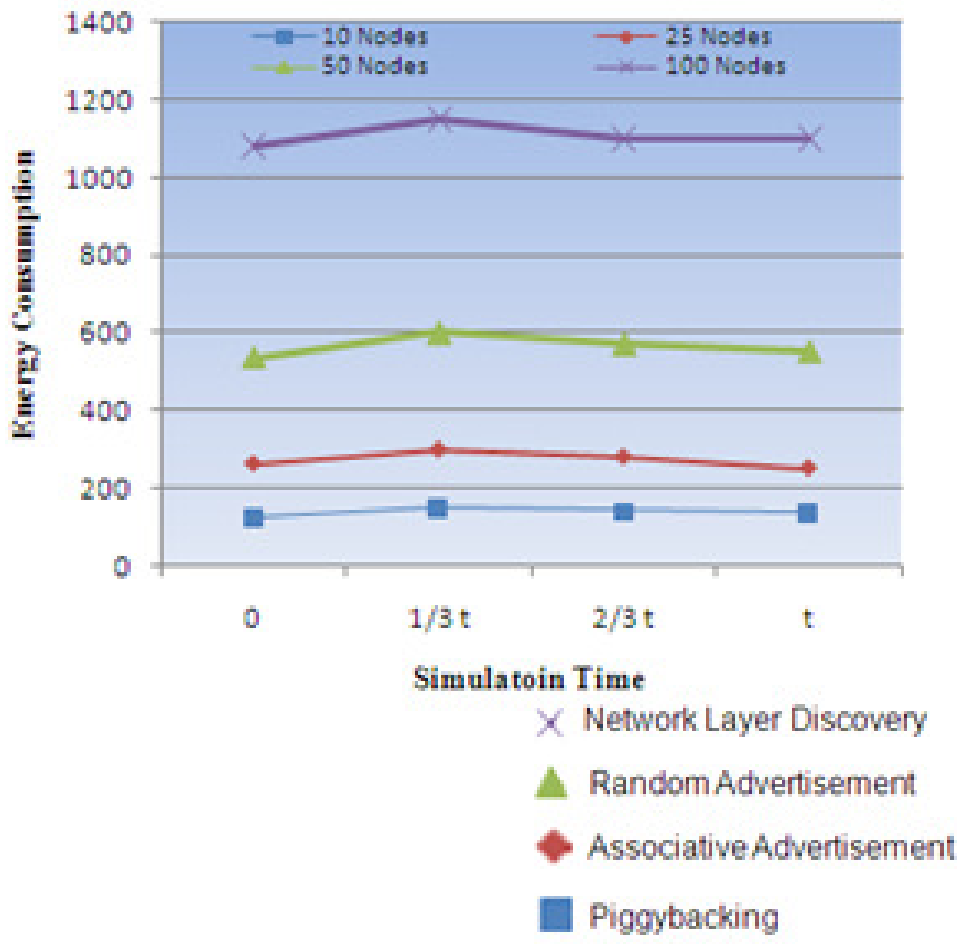

a) Energy consumption Vs # of nodes

b) Energy consumption by various components

| | | | | | |
|---|---|---|---|---|---|
| Number of Services | 64 | Correlation Matrix | YouTube | Cache Size | 8 |
| Routing Protocol | AODV | Mobility Model | Street Mobility | | |

**Figure 10: Analysis of energy consumption of Service Discovery Component**

Table 4 summarizes the results of the experiments performed. The improvement in latency, hit ratio and energy consumption with the proposed technique is shown. The total latency incurred to access a service is 20.3 sec with a hit ratio of 17.9% and energy consumption of 550KW.

**Table 4: Summary of experiments to analyze the service discovery component (for 50 nodes)**

| | Technique | Latency (sec) | Hit Ratio (%) | Energy Consumption (KW) |
|---|---|---|---|---|
| Reactive Discovery | | 22.5 | 8.2 | 535 |
| | Random Advertisement | -1.54 | +2.54 | -165 |
| | Associative Advertisement | -1.45 | +3.18 | +300 |
| | Piggybacking (FP-Growth) | -2.54 | +4 | 2+0 |
| | Piggybacking (Apriori [25]) | -2.27 | +3.99 | |
| | Semantic Processing | +3.3 | | |
| **Total** | | **20.3 sec** | **17.9 %** | **550 KW** |

### 6.2 Semantic Component

This section analyzes the semantic processing component of HANDY. As mentioned earlier, HANDY utilizes OWL-based multi-level ontology to address the semantic interoperability issue in a scalable manner. However, the semantic interoperability comes with processing overhead incurred to perform reasoning and inference engine. Fig 11 compares the latency when multilevel schema approach is adopted against the unified schema management approach. With multi-level approach, an additional overhead of 2.5 sec is observed. From the graph, it can however be observed that the multi-level approach yields improved performance (3.3 sec) as compared to the unified schema management approach.

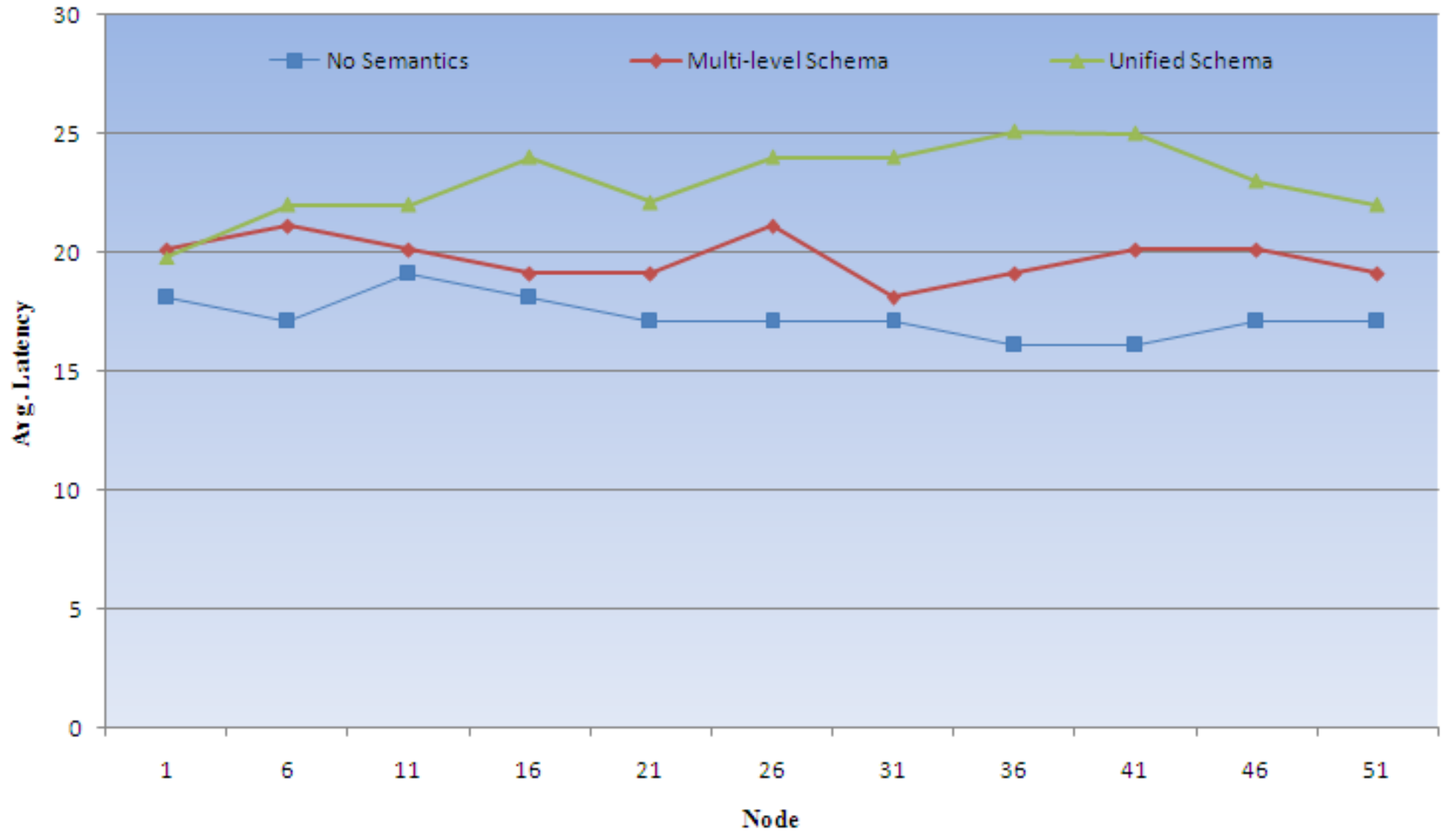

| | | | | | |
|---|---|---|---|---|---|
| Number of Nodes | 50 | Number of Services | 64 | Correlation Matrix | YouTube |
| Routing Protocol | DSR | Mobility Model | Street Mobility | Cache Size | 8 |

**Figure 11: Comparison of Latency with Semantic vs. Non-Semantic Discovery Component**

Fig 12 assesses the scalability of HANDY by comparing the performance of multilevel schema management with unified approach to service discovery. Fig 12a shows the distribution of ontology tuples at various stages of simulation on different network nodes. It can be observed that the ontology size grows gradually during simulation. Fig 12b analyzes the behavior of the proposed approach with different number of nodes by comparing the latency against the monolithic approach. The multi-level approach leads to graceful growth in latency as the number of nodes increases. However with the unified approach, the latency rises very sharply. Hence, it can be concluded that the proposed approach provides scalable discovery of services.

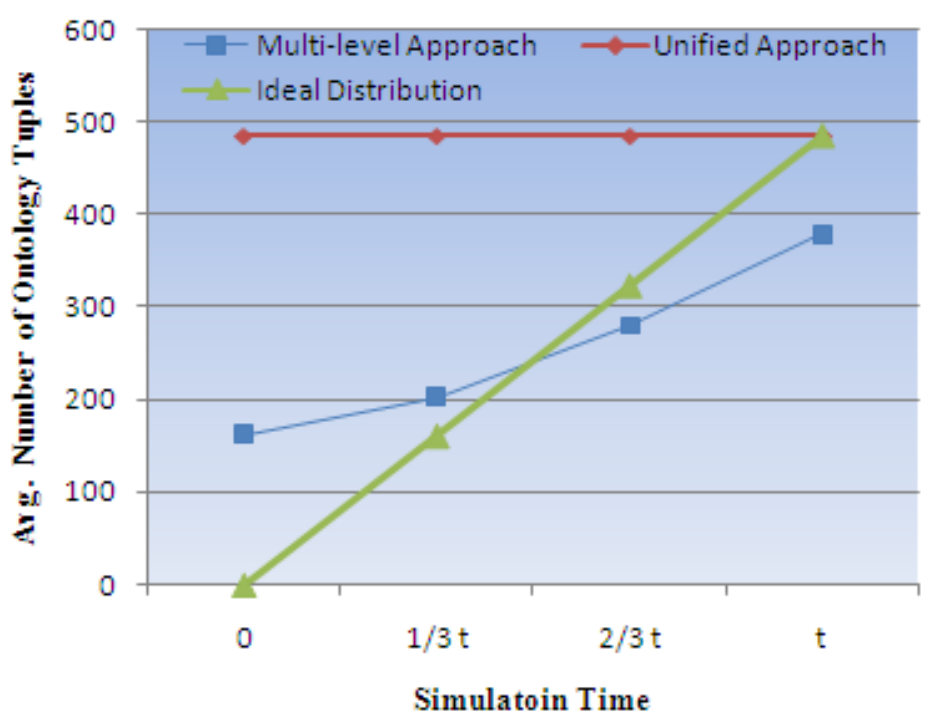

a) Ontology tuples during simulation (50 nodes)

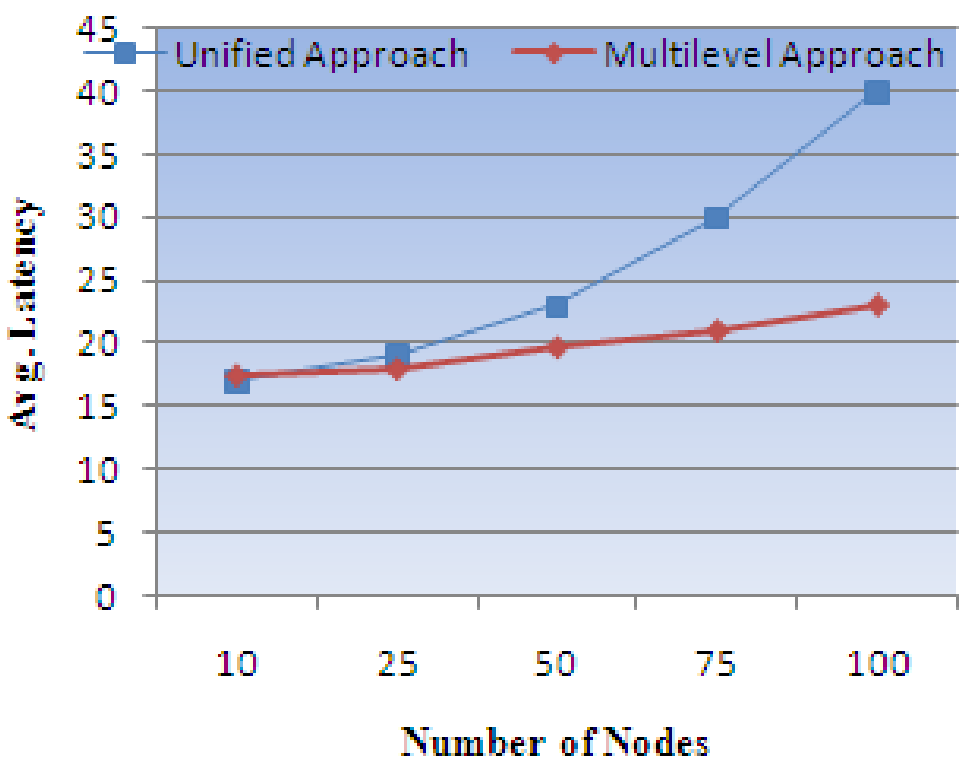

b) Latency with different number of nodes

| | | | | | |
|---|---|---|---|---|---|
| Number of Services | 64 | Correlation Matrix | YouTube | Cache Size | 8 |
| Routing Protocol | AODV | Mobility Model | Street Mobility | | |

**Figure 12: Scalability Analysis of Service Discovery Component**

# 7 Conclusion

In this paper, we have discussed a network layer semantic service discovery scheme HANDY that is based on exploiting the associations among services. The proposed scheme exploits different techniques to discover the services and the corresponding routing path simultaneously and then piggybacks responses of possible future discovery requests. A multilevel approach to maintain ontology is proposed that maintains fractional views of overall schema and discovers necessary ontology documents on demand. Through simulations, we have analyzed the hit ratio, latency and energy consumption in different network settings and concluded that the proposed discovery scheme gives rise to improved latency, hit ratio and scalability of the discovery process.

As far as the extensions to the current work are concerned, there are a multitude of tracks where the proposed work can be steered further. Further intelligence can be incorporated in the proactive component (HANDY-P) by using clever broadcasting of services based on other parameters like the available bandwidth, battery power, network congestion, service popularity etc. In this paper, we have provided the description of two reactive routing protocols AODV and DSR that can be used to for network layer discovery of the services on the network. However, there are other reactive routing protocols like TORA, ABR and OLSR etc. that can be explored for being used as the base routing protocol for the discovery of services. Approximate algorithms can be devised specifically targeted for MANET that minimizes the memory needs. In the proposed approach, FP-Growth Association Rules Mining algorithm has been employed as the data mining algorithm. However, this algorithm is not very efficient in terms of memory consumption. Future work can be done on development of new algorithms that minimize the memory needs. Similarly, in the proposed framework, the service cache maintained at individual nodes uses LRU for cache replacement. The utilization of the log database can be investigated for other operations. This includes service discovery planning, cache management and transaction management etc.

## Acknowledgement

This work has been supported by Higher Education Commission of Pakistan through its indigenous PhD program and Center for Research in Ubiquitous Computing (CRUC) at National University of Computer and Emerging Sciences (FAST-NU), Karachi, Pakistan.